\documentclass[english]{article}
\usepackage[OT4]{fontenc}
\usepackage[latin9]{inputenc}
\usepackage{float}
\usepackage{units}
\usepackage{textcomp}
\usepackage{amsmath}
\usepackage{graphicx}
\usepackage{amssymb}

\makeatletter
\newcommand{\lyxaddress}[1]{
\par {\raggedright #1
\vspace{1.4em}
\noindent\par}
}

\newtheorem{col}{Corollary}[section]
\newtheorem{twr}{Theorem}[section]

\makeatother

\begin{document}
\global\long\global\long\def\rr{\mathbb{R}}
 \global\long\global\long\def\cc{\mathbb{C}}
\global\long\global\long\def\tg{\tilde{\Gamma}}
 \global\long\def\todo#1{{\bf TODO: #1\/}}
\global\long\global\long\def\vin{V_{\Gamma}^{in}\left(k\right)}
\global\long\global\long\def\vout{V_{\Gamma}^{out}\left(k\right)}
\global\long\global\long\def\vtin{V_{\tg}^{in}}
\global\long\global\long\def\vtout{V_{\tg}^{out}}
\global\long\global\long\def\vv{\mathcal{V}}
 \global\long\def\ee{\mathcal{E}}
 \global\long\def\leads{\mathcal{L}}
\global\long\global\long\def\lgr{L_{\nicefrac{\tg}{R}}}
 \global\long\global\long\def\lgi{L_{\tg}^{(i)}}
\global\long\global\long\def\lgri{L_{\nicefrac{\tg}{R}}^{(i)}}

\title{Note on the role of symmetry in scattering from isospectral graphs
and drums}

\author{R Band$^{1,2}$%
\thanks{email: rami.band@bristol.ac.uk%
}~~A Sawicki$^{1,3}$%
\thanks{email: assawi@cft.edu.pl%
} and U Smilansky$^{2,4}$%
\thanks{email: uzy.smilansky@weizmann.ac.il%
}}

\maketitle

\lyxaddress{$^{1}$School of Mathematics, University of Bristol, University Walk,
Bristol BS8 1TW, UK }

\lyxaddress{$^{2}$Department of Physics of Complex Systems, The Weizmann Institute
of Science, Rehovot 76100, Israel}

\lyxaddress{$^{3}$Center for Theoretical Physics, Polish Academy of Sciences,
Al. Lotników 32/46, 02-668 Warszawa, Poland }

\lyxaddress{$^{4}$Cardiff School of Mathematics and WIMCS, Cardiff University,
Senghennydd Road, Cardiff CF24 4AG, UK }
\begin{abstract}
We discuss scattering from pairs of isospectral quantum graphs constructed
using the method described in \cite{BPB09,PB09}. It was shown in
\cite{BSS10} that scattering matrices of such graphs have the same
spectrum and polar structure, provided that infinite leads are attached
in a way which preserves the symmetry of isospectral construction.
In the current paper we compare this result with the conjecture put
forward by Okada $et$ $al.$ \cite{OSTH05} that the pole distribution
of scattering matrices in the exterior of isospectral domains in $\mathbb{R}^{2}$
are different.
\end{abstract}

\section{Introduction}

The examination of inverse spectral problems was initiated in 1966
by the famous question of Marc Kac 'Can one hear the shape of a drum?'
\cite{Kac66}. This question concerns the uniqueness of the spectrum
of Laplacian on planar domains with Dirichlet boundary conditions.
A substantial result towards answering Kac's question was due to Sunada
who presented a theorem that describes a method for constructing isospectral
Riemannian manifolds \cite{Sunada85}. In 1992, using an extension
of Sunada's theorem, Gordon, Webb and Wolpert answered Kac's question
as it related to drums, presenting the first pair of isospectral two-dimensional
planar domains \cite{GWW92a,GWW92b}. At the same time, the investigation
of scattering data started. Examples of objects that share the same
scattering information were found both for finite area \cite{Berard92,Zelditch92}
and infinite area Riemann surfaces \cite{GuiZwo97,Brooks02isoscatteringon}.
The search for isospectral and isoscattering examples now includes
objects ranging from Riemannian manifolds to discrete graphs. The
interested reader can find more about it in the reviews \cite{MR2179793,Gordon09,MR1705572}
and the references within.

The inverse spectral problem for quantum graphs has been first analyzed
by Gutkin and Smilansky \cite{GS01}. The authors of \cite{GS01}
proved that a simple graph with incommensurate lengths of the edges
can be fully reconstructed either from the spectrum of its Laplacian
or from the overall phase of its scattering matrix. In the recent
papers \cite{BPB09,PB09,BSS10} a construction method of isospectral
and isoscattering quantum graphs was presented. 

The work presented in this paper was originally motivated by a paper
of Okada $et$ $al.$ \cite{OSTH05}, in which the scattering from
the exterior of isospectral domains in $\mathbb{R}^{2}$ is discussed.
The authors suggest that, in spite of the fact that the two domains
are isospectral, when looked from exterior, the corresponding scattering
matrices are not isopolar. This proposition is not proved, but is
ushered by heuristic arguments based on the \char`\"{}interior - exterior\char`\"{}
duality, and augmented by numerical simulations. The simulations are
performed on the isospectral domains in $\mathbb{R}^{2}$ which were
constructed by Gordon $et$ $al.$ \cite{GWW92a,GWW92b} and on further
examples by Buser $et$ $al.$ \cite{BCDS94}. It is natural to test
this conjecture for graphs, which, in spite of being quite simple,
usually display most of the complex features which characterize the
spectra of domains in $\mathbb{R}^{2}$ studied as interior or exterior
problems. Following \cite{BSS10} we prove that for every pair of
isospectral quantum graphs obtained from the construction presented
in \cite{BPB09,PB09} the scattering matrices have the same polar
structure, provided that leads are attached in a way which preserves
the symmetry of isospectral construction. We call such graphs isoscattering.
We compare this result with the conjecture put forward by Okada $et$
$al.$ and explain that there is no conflict between these two results.
The general proofs of most of the statements can be found in \cite{BSS10}.
In the current paper we present some examples and emphasize the role
of the symmetry in this problem.

\section{Quantum graphs and scattering matrices}

\subsection{Quantum graphs}

\label{sec:quantum_graphs}

Let $\Gamma=\left(V,\, E\right)$ be a finite graph which consists
of $\left|V\right|$ vertices that are connected by $\left|E\right|$
edges. Each edge, $e\in E$, is a one dimensional segment of finite
length $L_{e}$ with a coordinate $x_{e}\in\left[0,L_{e}\right]$
and this makes $\Gamma$ a metric graph. The metric graph becomes
quantum, when we supply it with a differential operator. Here we choose
our operator to be the free Schrödinger operator and denote it by
$\Delta$. This is merely the one-dimensional Laplacian which equals
$-\frac{{\rm d}^{2}}{{\rm d}x_{e}^{2}}$ on each of the edges $e\in E$.
The coupling between the edges is introduced by vertex conditions
at the vertices. In the following we will only use Neumann and Dirichlet
vertex conditions (see \cite{GS06,Kuchment04} for other possibilities),
which are described below. Let $v\in V$, and $E_{v}$ the set of
edges incident to $v$. A function $f$ on $\Gamma$ obeys Neumann
vertex conditions at $v$ if and only if 
\begin{enumerate}
\item $f$ is continuous at $v$, i.e., \[
\forall e_{1},e_{2}\in E_{v}\,;\,\, f_{e_{1}}\left(v\right)=f_{e_{2}}\left(v\right).\]

\item The sum of derivatives of $f$ at the vertex $v$ equals zero.\[
\sum_{e\in E_{v}}\frac{{\rm \textrm{d}f}}{\textrm{d}x_{e}}\left(v\right)=0.\]

\end{enumerate}
A function $f$ on $\Gamma$ obeys Dirichlet vertex conditions at
$v$ if and only if\[
\forall e\in E_{v}\,;\,\, f_{e}\left(v\right)=0,\]
and there are no further requirements on the derivatives.

\subsection{The scattering matrix of a quantum graph}

In this section we explain how to define the scattering matrix of
a quantum graph. In order to speak about the scattering problem for
a quantum graph we need to connect its vertices (all or a subset)
by leads which extend to infinity. For simplicity we assume that leads
are connected to vertices of valency grater or equal than two supplied
with Neumann vertex conditions. Let us denote by $\tg$ the extended
quantum graph which consists of the original graph $\Gamma$ and the
external leads $\mathcal{L}$ connected to $M\leq|V|$ vertices which
we call the marked vertices. We will not elaborate here on the vertex
conditions for $\tilde{\Gamma}$ (see \cite{BBS10} for a more detailed
discussion). In the following we will use the two rules 
\begin{itemize}
\item the non-marked vertices are supplied with the same vertex conditions
as they had in $\Gamma$
\item at each marked vertex $v$ we have Neumann vertex conditions. 
\end{itemize}
We now introduce the scattering matrix which corresponds to $\tg$
and denote it by $S_{\tg}$. 

\noindent Let $f$ be an eigenfunction of $\Delta$ with eigenvalue
$k^{2}$ and let $\leads$ be the set of leads connected to $\Gamma$.
The restriction of $f$ to the lead $l\in\leads$ can be written in
the form

\begin{equation}
f_{l}\left(x_{l}\right)=a_{l}^{\, in}\exp\left(-ikx_{l}\right)+a_{l}^{\, out}\exp\left(ikx_{l}\right).\label{eq:function_on_leads}\end{equation}
Collecting all the variables $\left\{ a_{l}^{\, in}\right\} _{l\in\leads}$
and $\left\{ a_{l}^{\, out}\right\} _{l\in\leads}$ into vectors which
we denote by $\vec{a}^{\, in}$ and $\vec{a}^{\, out}$, we introduce
the shorthand notation\begin{equation}
\left.f\right|_{\leads}=\vec{a}^{\, in}\exp\left(-ikx\right)+\vec{a}^{\, out}\exp\left(ikx\right).\end{equation}
Using the requirements dictated by the vertex conditions on all the
vertices of the graph $\tilde{\Gamma}$, we may write a set of linear
equations, some of whose variables are $\left\{ a_{l}^{\, in}\right\} _{l\in\leads}$
and $\left\{ a_{l}^{\, out}\right\} _{l\in\leads}$. Solving these
equations yields relation\begin{equation}
\vec{a}^{\, out}=S_{\tg}\left(k\right)\vec{a}^{\, in}.\label{eq:basic_scattering_relation}\end{equation}
The matrix $S_{\tg}\left(k\right)$ is a square matrix of dimension
$\left|\leads\right|$ and is unitary for every $k\in\rr$. This is
the scattering matrix of the graph $\tg$ (also called the scattering
matrix of $\Gamma$). The existence and uniqueness of $S_{\tg}\left(k\right)$
for every value of $k$ and the unitarity of it on the real axis are
proved in \cite{BBS10}.

\section{Isospectral graphs\label{sec:isospectrality_and_transplantation}}

A new construction method of isospectral objects has been recently
presented in \cite{BPB09,PB09}. It is a generalization of the well-known
Suanda's construction \cite{Sunada85}. This method can be applied
to any geometric object. However here we bring the relevant aspects
of the theory as it applies to quantum graphs. In order to avoid quite
abstract formalism of representation theory we present the underlying
idea on the one particular example (the full discussion of this example,
as well as many others can be found in \cite{BPB09}).

\subsection{An Example\label{sub:An-Example}}

Let us consider the graph $\Gamma$ given in fi{}gure 1(a), where
$a,\, b,\, c$ are lengths of the edges and the vertex conditions
at all vertices are Neumann. All the symmetries of $\Gamma$ form
a group which is the dihedral group $G=D_{4}$ and it is group of
the symmetries of the square. Let us examine two subgroups of $G$

\begin{equation}
H_{1}=\{e,\, r_{u},\, r_{v},\,\sigma^{2}\},\quad H_{2}=\{e,\, r_{x},\, r_{y},\,\sigma^{2}\},\label{eq:H1H2}\end{equation}
where $r_{x},\, r_{y},\, r_{u},\, r_{v}$ denote reflections by the
axes $x,\, y,\, u,\, v$ and $\sigma$ is the counterclockwise rotation
by $\nicefrac{\pi}{2}$. Consider the following one dimensional representations
$R_{1}$ and $R_{2}$ of $H_{1}$ and $H_{2}$ respectively

\begin{gather}
R_{1}=\{e\text{\textrightarrow}(1),\,\text{\ensuremath{\sigma}\textrightarrow}(\text{\textminus}1),\, r_{v}\text{\textrightarrow}(-1),\, r_{u}\text{\textrightarrow}(1)\},\label{eq:R1R2}\\
R_{2}=\{e\text{\textrightarrow}(1),\text{\,\ensuremath{\sigma}\textrightarrow}(\text{\textminus}1),\, r_{y}\text{\textrightarrow}(1),\, r_{x}\text{\textrightarrow}(\text{\textminus}1)\}.\nonumber \end{gather}
Using these representations we will construct two graphs denoted by
$\nicefrac{\Gamma}{R_{1}}$ and $\nicefrac{\Gamma}{R_{2}}$ (fi{}gure
1(b), 1(c)) which are isospectral. Now we will explain the process
of building the quotient graph $\nicefrac{\Gamma}{R_{2}}$. To this
end assume that $f$ is the eigenfunction of the Schrödinger operator
on the graph $\Gamma$ with eigenvalue $k^{2}\in\mathbb{R}$ which
transforms according to the representation $R_{2}$, i.e.

\begin{gather}
\forall g\in H_{2}\,\,\,\, gf=R_{2}(g)f,\end{gather}
where the action of $G$ on $f$ is

\begin{gather}
[gf](x)=f(g^{-1}x),\end{gather}
and $R_{2}(g)$ is specified in (\ref{eq:R1R2}). Since the function
$f$ transforms according to the representation $R_{2}$ we know that
$r_{x}f=-f$ . This implies that $f$ is anti-symmetric with respect
to the horizontal refl{}ection and in particular vanishes on the fi{}xed
points of $r_{x}$. Similarly $r_{y}f=f$ which means that $f$ is
symmetric with respect to the vertical refl{}ection, hence the derivative
of $f$ vanishes at the fi{}xed points of $r_{y}$. Notice now that
it is enough to know $f$ on the graph shown in figure 1(c) in order
to deduce $f$ on the whole graph. The graph shown in figure 1(c)
is called the quotient graph $\nicefrac{\Gamma}{R_{2}}$. Repeating
the same procedure for $R_{1}$ we obtain the graph $\nicefrac{\Gamma}{R_{1}}$
(see figure 1(b)). 

\begin{figure}[H]
\includegraphics[scale=0.3]{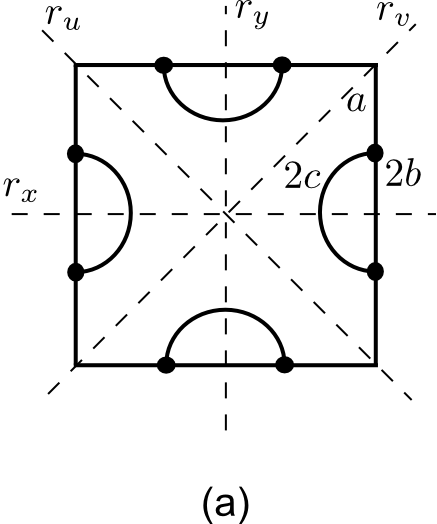}\qquad{}\qquad{}~~~~~~\includegraphics[scale=0.3]{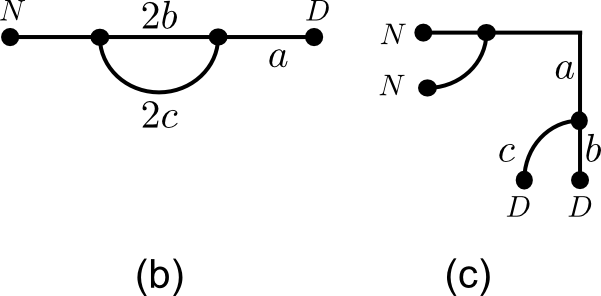}

\caption{(a) The graph $\Gamma$ that obeys the dihedral symmetry of the square.
The lengths of some edges and the axes of the reflection elements
in $D_{4}$ are marked, (b) The graph $\nicefrac{\Gamma}{R_{1}}$,
(c) The graph $\nicefrac{\Gamma}{R_{2}}$}
\label{Flo:graphgamma}
\end{figure}
It turns out that graphs $\nicefrac{\Gamma}{R_{1}}$ and $\nicefrac{\Gamma}{R_{2}}$
are isospectral \cite{BPB09,PB09}. Moreover the isospectrality of
these graphs is due to the fact that \begin{gather}
\mathrm{Ind}_{H_{1}}^{G}R_{1}\simeq\mathrm{Ind}_{H_{2}}^{G}R_{2},\label{eq:ind}\end{gather}
and the construction method described above.

\subsection{A transplantation\label{sub:A-transplantation}}

Let us denote by $\Phi_{\Gamma}(k)$ the eigenspace of $\Delta$ corresponding
to eigenvalue $k^{2}$. In this section we will explain the concept
of \emph{transplantation}. The transplantation is a map between isospectral
graphs 

\begin{equation}
T:\Phi_{\Gamma_{1}}\left(k\right)\overset{\cong}{\longrightarrow}\Phi_{\Gamma_{2}}\left(k\right),\end{equation}
which assigns to every eigenfunction on $\Gamma_{1}$ with eigenvalue
$k^{2}$ an eigenfunction on $\Gamma_{2}$ with the same eigenvalue
$k^{2}$. The way transplantation acts can be easily understood. Notice
that from the figure 1(b), 1(c) we see that the isospectral objects
consist of some elementary \char`\"{}building blocks\char`\"{} that
are attached to each other in two different prescribed ways. The transplantation
can be usually expressed in terms of these building blocks. It expresses
the restriction of an eigenfunction to a building block of the first
object as a linear combination of the restrictions of an eigenfunction
on building blocks of the second object. In case of the graphs in
figure 1(a), 1(b) each of them consists of two building blocks and
the transplantation matrix from $\nicefrac{\Gamma}{R_{1}}$ to $\nicefrac{\Gamma}{R_{2}}$
is given by 

\begin{gather}
T=\left(\begin{array}{cc}
1 & 1\\
1 & -1\end{array}\right).\label{eq:T}\end{gather}
It is worth to mention that the existence of transplantation in this
case is not a surprise, i.e., the isospectral construction method
described in \cite{BPB09,PB09} always yields a transplantation.

\section{Isoscattering graphs and drums \label{sec:examples}}

\subsection{Isoscattering graphs\label{sub:Isoscattering-graphs}}

In this section we give an example of isoscattering graphs, i.e.,
graphs for which scattering matrices have the same poles structure.
The full discussion of a recently presented construction method of
isoscattering graphs can be found in \cite{BSS10}. 

In section \ref{sub:An-Example} we explained how to construct isospectral
graphs. The method was mainly due to the symmetry of the parent graph
$\Gamma$. It is easy to see that we can repeat this procedure for
a graph $\tilde{\Gamma}$ with leads (see figure 2). 

\begin{figure}[H]
\includegraphics[scale=0.3]{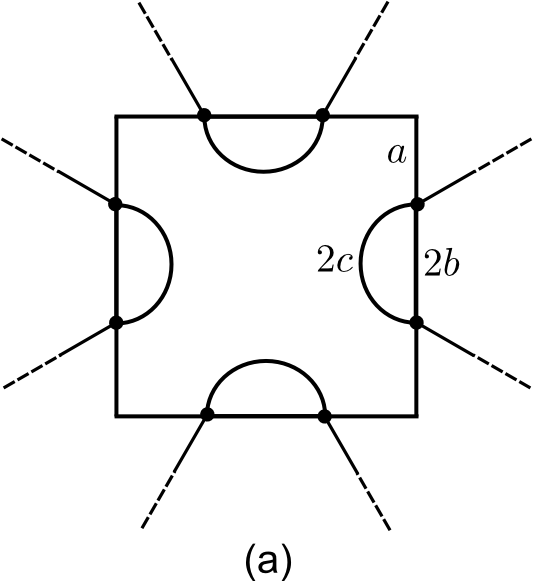}\qquad{}\qquad{}~~~~~~\includegraphics[scale=0.3]{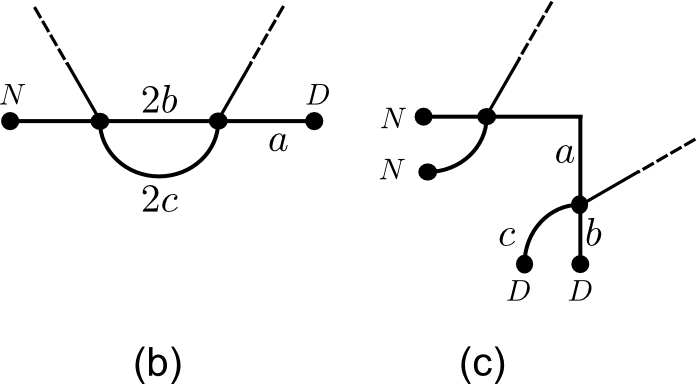}\caption{(a) The Graph $\tilde{\Gamma}$ with leads attached, (b) The graph
$\nicefrac{\tilde{\Gamma}}{R_{1}}$, (c) The graph $\nicefrac{\tilde{\Gamma}}{R_{2}}$}
\label{Flo:exten}
\end{figure}

\noindent Moreover, there is a transplantation $T$ between graphs
$\nicefrac{\tilde{\Gamma}}{R_{1}}$ and $\nicefrac{\tilde{\Gamma}}{R_{2}}$
and it is given by (\ref{eq:T}). We know from section \ref{sub:A-transplantation}
that the transplantation is a linear transformation which sends every
eigenfunction from $\nicefrac{\tilde{\Gamma}}{R_{1}}$ to an eigenfunction
of $\nicefrac{\tilde{\Gamma}}{R_{2}}$. The following two observations
turn out to be of great importance
\begin{enumerate}
\item It is possible to restrict the transplantation to the leads. This
is since a function restricted to a lead of $\nicefrac{\tilde{\Gamma}}{R_{1}}$
can be only send to some linear combination of a function restricted
to leads of $\nicefrac{\tilde{\Gamma}}{R_{2}}$ and vice verse. 
\item For the graphs $\nicefrac{\tilde{\Gamma}}{R_{i}}$ the restriction
of an eigenfunction to the leads is of the form \begin{equation}
\left.f_{\nicefrac{\tilde{\Gamma}}{R_{i}}}\right|_{\leads}=\vec{a}_{\nicefrac{\tilde{\Gamma}}{R_{i}}}^{\, in}\exp\left(-ikx\right)+S_{\nicefrac{\tilde{\Gamma}}{R_{i}}}(k)\vec{a}_{\nicefrac{\tilde{\Gamma}}{R_{i}}}^{\, in}\exp\left(ikx\right),\end{equation}
 where $S_{\nicefrac{\tilde{\Gamma}}{R_{i}}}(k)$ is the corresponding
scattering matrix, $\vec{a}^{\, in}\in\mathbb{C}^{|\mathcal{L}|}$.
\end{enumerate}
The existence of a transplantation together with observation 1 gives
\begin{eqnarray}
\left.f_{\nicefrac{\tilde{\Gamma}}{R_{1}}}\right|_{\mathcal{L}} & = & \vec{a}_{\nicefrac{\tilde{\Gamma}}{R_{1}}}^{\, in}\exp\left(-ikx\right)+S_{\nicefrac{\tilde{\Gamma}}{R_{1}}}(k)\vec{a}_{\nicefrac{\tilde{\Gamma}}{R_{1}}}^{\, in}\exp\left(ikx\right)\\
 & \downarrow T\nonumber \\
\left.f_{\nicefrac{\tilde{\Gamma}}{R_{2}}}\right|_{\mathcal{L}} & = & \vec{a}_{\nicefrac{\tilde{\Gamma}}{R_{2}}}^{\, in}\exp\left(-ikx\right)+S_{\nicefrac{\tilde{\Gamma}}{R_{2}}}(k)\vec{a}_{\nicefrac{\tilde{\Gamma}}{R_{2}}}^{\, in}\exp\left(ikx\right).\nonumber \end{eqnarray}
So we obtain \begin{gather}
\vec{a}_{\nicefrac{\tilde{\Gamma}}{R_{2}}}^{\, in}=T\vec{a}_{\nicefrac{\tilde{\Gamma}}{R_{1}}}^{\, in}\\
S_{\nicefrac{\tilde{\Gamma}}{R_{2}}}(k)\vec{a}_{\nicefrac{\tilde{\Gamma}}{R_{2}}}^{\, in}=TS_{\nicefrac{\tilde{\Gamma}}{R_{1}}}(k)\vec{a}_{\nicefrac{\tilde{\Gamma}}{R_{1}}}^{\, in}.\nonumber \end{gather}
 Finally we get \begin{equation}
S_{\nicefrac{\tilde{\Gamma}}{R_{2}}}T\vec{a}_{\nicefrac{\tilde{\Gamma}}{R_{1}}}^{\, in}=TS_{\nicefrac{\tilde{\Gamma}}{R_{1}}}(k)\vec{a}_{\nicefrac{\tilde{\Gamma}}{R_{1}}}^{\, in}\,\,\,\Rightarrow\,\,\, T^{-1}S_{\nicefrac{\tilde{\Gamma}}{R_{2}}}(k)T=S_{\nicefrac{\tilde{\Gamma}}{R_{1}}}(k).\end{equation}
The following are now justified 

\begin{twr} The scattering matrices of $\nicefrac{\tilde{\Gamma}}{R_{1}}$
and $\nicefrac{\tilde{\Gamma}}{R_{2}}$ are conjugated by the transplantation
map for every $k\in\mathbb{C}$.

\end{twr} 

\begin{col} The scattering matrices of $\nicefrac{\tilde{\Gamma}}{R_{1}}$
and $\nicefrac{\tilde{\Gamma}}{R_{2}}$ have the same polar structure.

\end{col} 

\noindent This way we get that two graphs $\nicefrac{\tilde{\Gamma}}{R_{1}}$
and $\nicefrac{\tilde{\Gamma}}{R_{2}}$ are isoscattering. It is important
to notice that the construction of isoscattering graphs involve the
following ingredients/steps
\begin{itemize}
\item A graph $\Gamma$ with certain symmetry group
\item Two isospectral quotient graphs $\nicefrac{\Gamma}{R_{1}}$ and $\nicefrac{\Gamma}{R_{2}}$
\item A graph $\tilde{\Gamma}$ which is an extension of $\Gamma$ by attaching
leads to infinity in such a way that the graph $\tilde{\Gamma}$ obeys
the same symmetry group as $\Gamma$
\item Two isoscattering quotient graphs $\nicefrac{\tilde{\Gamma}}{R_{1}}$
and $\nicefrac{\tilde{\Gamma}}{R_{2}}$.
\end{itemize}

\subsection{Scattering from isospectral drums}

In this section we go back to the the conjecture put forward by Okada
$et$ $al.$ stating that the pole distribution of scattering matrices
in the exterior of isospectral domains in $\mathbb{R}^{2}$ are different.
At first glance this conjecture seems to be in contradiction with
the result presented in section \ref{sub:Isoscattering-graphs}. Our
main goal is to understand that there is no conflict here.

\noindent Let us first consider the two isospectral drums presented
in figure 3. Their isospectrality was first proved by Gordon $et$
$al.$ \cite{GWW92a}. In \cite{OSTH05} the scattering problem for
these two drums was investigated. In particular, the poles structure
of scattering matrices of these two drums were computed numerically.
The authors of \cite{OSTH05} found that these structures are different
and hence they concluded that it is possible to distinguish between
these two drums while looking from the outside.

\begin{figure}[H]
~~~~~~\includegraphics[scale=0.6]{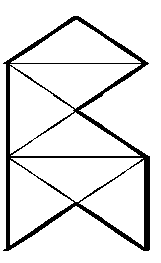}~~~~~~~~~~~~~~~~~~~~~~~~~~~~~~~~~~~~~~~~~~~~~~\includegraphics[scale=0.6]{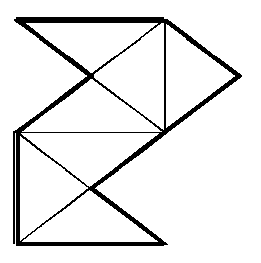}

\caption{Two isospectral drums.}

\end{figure}

\noindent It was first noticed by Buser \cite{BCDS94} that the isospectrality
of these two drums can be proved using Sunada's construction applied
to hyperbolic plane. Sunada's method involves all the elements used
in the recently presented method \cite{BPB09,PB09} albeit the algebraic
condition (\ref{eq:ind}) is restricted to the trivial representations

\begin{gather}
\mathrm{Ind}_{H_{1}}^{G}\mathrm{id}\simeq\mathrm{Ind}_{H_{2}}^{G}\mathrm{id}.\label{eq:inddrum}\end{gather}
In particular the construction of quotient is analogous to the one
described in section \ref{sub:An-Example}. We will now describe this
construction. To this end we treat the hyperbolic plane (see figure
4) as a 'graph' with symmetries - each of the lines in figure 4 represents
one reflection symmetry. Desymmetrization of the hyperbolic plane
along some particular reflection lines yields two compact domains
denoted in green in figure 4. Since our choice of reflection subgroups
fulfills the algebraic condition (\ref{eq:inddrum}) we get that these
two domains are isospectral. The immediate consequence of this construction
is that the isospectral hyperbolic drums are isometric, hence isoscattering.
Since their isospectrality is governed just by the construction method
we can replace the hyperbolic triangles by the Eucliden ones and this
way obtain the two isospectral drums shown in figure 3. Obviously,
the symmetry of the hyperbolic plane is no longer present for these
drums. Let us now consider the scattering from the drums shown in
figure 3. Going back to our graph analogy it is similar to attaching
infinite leads to graphs $\nicefrac{\Gamma}{R_{1}}$ and $\nicefrac{\Gamma}{R_{2}}$
in a way which does not come from the original symmetry of $\Gamma$.
Then of course the scattering matrices of the corresponding quotient
graphs have no longer the same polar structure. In case of drums the
same phenomena is present. In order to have the same poles of the
scattering matrices we need to consider the scattering problem on
the hyperbolic plane. Summing up there is no conflict between our
result for quantum graphs and the result of Okada $et$ $al.$ for
drums. Moreover the symmetry arguments are responsible for both of
them. 

\begin{figure}[H]
~~~~~~\includegraphics[scale=0.6]{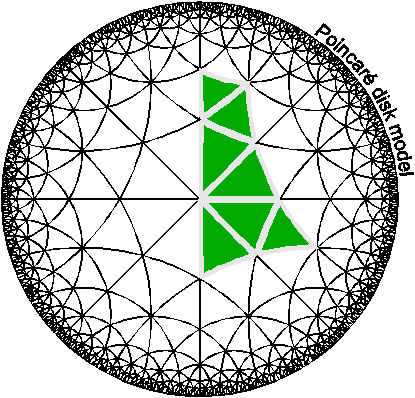}~~~~~~~~~~~\includegraphics[scale=0.6]{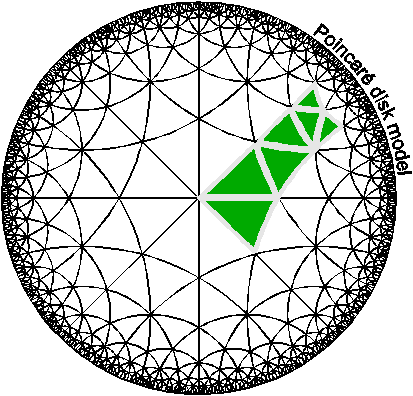}

\caption{The hyperbolic plane together with two isospectral hyperbolic drums.}

\end{figure}

\section{Acknowledgments}

We thank Akira Shudo and Ori Parzanchevski for fruitful discussions.
A Sawicki would like to thank the organizers of 5th Workshop on Quantum
Chaos and Localization Phenomena 20 - 22 May 2011 - Warsaw, Poland
for their invitation and hospitality. Grants from EPSRC (grant EP/G021287),
ISF (grant 166/09) and BSF (grant 2006065) are acknowledged. The support
by Polish MNiSW grant no. N N202 085840 is gratefully acknowledged.
R Band is supported by EPSRC grant no. EP/H028803/1.

\end{document}